\documentstyle [12pt] {article} \textwidth 160mm\textheight
220mm \topmargin - 2mm \oddsidemargin -5mm\evensidemargin -5mm
\begin{document}

\begin {center} {\Large \bf The energy conservation law in classical
electrodynamics} \end {center}
\begin {center} {\large E.G.Bessonov} \end {center}

              \begin{abstract}
In the framework of the classical Maxwell-Lorentz electrodynamics the
energy conservation law is reconsidered.
              \end{abstract}

                       \section {Introduction}

The Poynting theorem has been a corner-stone of electromagnetic theory
since its publication in 1884. Its success have been so many that its
limitations have been denied. The classical electrodynamics is
considered to be the consistent relativistic theory. It was emphasized
that some difficulties which appear e.g. when the concept of particles
is introduced or when the equation of motion of particles in view of
radiation reaction force is derived are non-principal. In addition
everybody refers to the Poynting theorem that is to the laws of
conservation of energy, linear and angular momentum of the combined
system of particles and fields.

In the present paper we pay attention to a typical logical  mistake
made by founders of classical electrodynamics and adopted by repetition
by authors of textbooks on classical electrodynamics in the case when
the law of conservation of energy is derived for a system consisting of
electromagnetic field and charged particles. The violation of the law
is displayed when the energy of particles of the system is changed.  It
means that this law should be treated as a new open question of
classical electrodynamics and references to this law are incorrect when
any difficulties are discussed.

Below we will present the detailed and typical proof of the Poynting's
theorem in the framework of Maxwell electrodynamics and the law of the
energy conservation in the electrodynamics of Maxwell-Lorentz for a
system consisting of an electromagnetic field and charged particles.
Then we will specify a logic error in the last proof.

\section {The laws of conservation in electrodynamics}

Let us consider the matter of charged particle is a continuous media.
Then from the Maxwell equations an equation follows

            \begin{equation} 
            {\partial \over \partial t} \int _V{|\vec
            E | ^2 + | \vec H | ^2\over 8\pi} dV + \oint \vec S\, d\vec
            f = - \int _V\vec \jmath \vec E dV \end{equation}
where $\vec E$, $\vec H$ are vectors of an electric and magnetic field
strengths, $\vec \jmath = \rho \vec v$ vector of a current density,
$\rho$ a charge density, $\vec v$ vector of velocity of motion of the
given element of volume of the charge,  $\vec S = (c/4\pi) [\vec E\vec
H] $ the Pointing vector, $c$ a velocity of light, $d\,\vec f = \vec
n\,df$, $\vec n$ the unit vector normal to the surface limiting some
volume $V$, $df$ the element of the area of the surface (see Appendix
1) \cite{landau}.  The value $w = {| \vec E | ^2 + | \vec H | ^2/
8\pi}$ in the Eq(1) is the density of the energy of the electromagnetic
field.  If the integration in the Eq(1) will be made over the whole
infinite space than the term with the Poynting's vector can be omitted.
In this case the electromagnetic fields emitted by particles in the
form of electromagnetic waves will not go out of the space.

The Eq(1) is derived from the microscopic Maxwell equations. If we
replace in the Eq(1) the values $| \vec E | ^2$ and $| \vec H | ^2$ by
$| \vec E ||\vec D|$ and $| \vec H ||\vec B|$, where $|\vec D|$ and
$|\vec B|$ are the vectors of the electric and magnetic inductions then
we will have the modified equation for the macroscopic electrodynamics
\cite{jack}. In this case the modified equation is the statement of
conservation of energy or the well-known Poynting's theorem written in
the integral form: the time rate of change of electromagnetic energy
within a certain volume plus the energy flowing out through the
boundary surfaces of the volume per unit time, is equal to the negative
of the total work per unit time $dQ /dt = \int _V\vec \jmath \vec
E\,dV$ done by the fields on the sources within the volume.  The energy
$\varepsilon ^{nem} = \int _{-\infty} ^{t}(dQ/dt) dt$ can be treated
through the non-electromagnetic energy liberated in the volume $V$ by
the exited currents (heat energy or mechanical energy of neutral atoms
movement in the heated material bodies). It does not include any energy
of the electromagnetic origin. All electromagnetic energy is included
in the first term.

The Pointing's theorem based on the Maxwell equations and the
definition of the work done by the electric fields on currents had led
to the concept of the energy density of the electromagnetic field $w$
and to the concept of the Poynting vector $\vec S$ representing the
energy flow. It emphasised the consistency of the Maxwell
electrodynamics.

Later the Poynting theorem was generalized in reference to the case of
a combined system of particles and fields. A set of Maxwell equations
for the electromagnetic fields and Lorentz equations for the particles
were used. The way of generalization we can get e.g. from the textbook
\cite{jack}:  "The emphasis so far has been on the energy of the
electromagnetic fields. The work done per unit time per unit volume by
the fields ($\vec \jmath \cdot \vec E$) is a conversion of
electromagnetic into mechanical or heat energy. Since matter is
ultimately composed of charged particles (electrons and atomic nuclei),
we can think of this rate of conversion as a rate of the charged
particles energy increase per unit volume. Then we can interpret
Poynting's theorem for the microscopic fields ($\vec E, \vec H)$ as a
statement of the energy conservation in the combined system of
particles and fields. If we denote the total energy of the particles
within the volume $V$ as $\varepsilon ^{mech}$ and assume that no
particles move out of the volume, we have $d\varepsilon ^{mech}/dt =
\int _V (\vec \jmath \cdot \vec E)dV$. Then Poynting's theorem
expresses the conservation of energy for the combined system as

            \begin {equation} 
            {d\over dt}(\varepsilon ^{mech}
            + \varepsilon ^{em}) = - \oint _S \vec Sd\vec f.
            \end {equation}
where the total energy of the electromagnetic fields within the volume
$V$ is

            \begin {equation}        
            \varepsilon ^{em} = \int _Vw\,dV = {1\over 8\pi}\int
            _V({|\vec E|}^2 + {|\vec H|}^2)\,dV." \end {equation}

The value $\varepsilon ^{mech}$ is the sum of energies of particles
$\varepsilon _i$ and the value $ \varepsilon
^{em}$ the total energy of the electromagnetic field in the volume $V$.
If the volume $V$ is infinitely large then $\oint _S(\vec n\cdot \vec
S) = 0$ (the emitted energy never reach the boundary surface of the
volume $V$) and the expression (2) can be written in the form
$d\varepsilon _{\Sigma}/dt = 0$ or \cite{landau} \footnote{We
refer to the most popular textbooks \cite{landau}, \cite{jack}. One can
see that another textbooks both written long ago or recently have the
same presentation of this topic.}

             \begin{equation} 
             \varepsilon _{\Sigma} = \sum _i  \varepsilon _i +
             {1\over 8\pi}\int_V({|\vec E|}^2 + {|\vec H|}^2)dV =
             const.  \end{equation}

According to (4) the total energy $\varepsilon _{\Sigma}$ of the
combined system of particles and fields in the whole space is constant.
This expression is treated as the integral form of the law of
conservation of energy in the electrodynamics of Maxwell-Lorentz for
the case when the system is located in the whole space. This treatment
has a logical error and will be discussed below.

The first term in (4) is the sum of the energies of particles
$\varepsilon _i$. It is considered that the energy of a particle is the
value $\varepsilon _i = m_ic^2 \gamma _i$, where $m_i$ is a mass,
$\gamma _i = 1/\sqrt {1 - v_i^2/c^2}$ is the relativistic factor and
$v_i$ a velocity of a particle. The dependence of the particle's energy
on velocity is determined by the requirements of the special theory of
relativity. At introducing of the concept of particles in
electrodynamics it is usually postulated, that the energy of a particle
consists partially of the self-energy (or "inertial energy") of the
electromagnetic origin $\varepsilon _i^{em} $ and partially from the
field energy of the non-electromagnetic origin $\varepsilon _i^ {nem} $
($\varepsilon _i = \varepsilon _i^ {em} + \varepsilon _i^ {nem} $)
\cite{landau}.

Vectors $\vec E$, $\vec H$ in (4) represent the total electromagnetic
field created by a system of particles and independent fields.  Volume
integral of the energy density of the electromagnetic field in the
second term of the expression (4) represents the electromagnetic energy
of the system.  In the simplest case, when the independent fields
are absent the electromagnetic field strengths can be presented in the
form $\vec E = \sum \vec E_i$, $\vec H = \sum \vec H_i$ and the
electromagnetic energy of the system can be presented in the form
$\varepsilon ^{em} = \sum W ^{em}_i + \sum W ^{em}_{ij}$, where the
energy $W^{em}_i = (1/8\pi)\int (|\vec E_i|^2 + |\vec H_i|^2)dV$
corresponds to the energy of fields produced by a particle $i$ and the
energy $W^{em}_{ij} = (1/4\pi)\int (|\vec E_i\vec E_j| + |\vec H_i\vec
H_j|)dV$ corresponds to the mutual energy (generalized "interaction
energy" or "potential energy") of the electromagnetic fields.

After these remarks we can see that the energy of particles included in
the expression (4) twice in a different forms: the mass term $\sum
\varepsilon _i$ represents the total energy of particles and the field
term $\sum W ^{em}_i$ includes the energy of particles of the
electromagnetic origin (after the interaction when the particles will
move with a constant velocity for a long time the value $W ^{em}_i =
\varepsilon _i^{em}$)\footnote{In general case the energy
$W_i^{em}$ includes both the self-energy $\varepsilon_i^{em}$  and the
energy of the spontaneous incoherent radiation of the particle which
can be selected only in a wave zone. The generalized "interaction
energy" $W^{em}_{ij}$ have sense of the potential energy only in
statics.}. From the other hand if we will treat the total energy of a
particle as the sum of energies of non-electromagnetic and
electromagnetic origin \cite{landau} then it will mean that the energy
of particles of the electromagnetic origin will be included in the
expression (4) twice. All that means that the standard treatment of the
expression (4) is incorrect.\footnote {The expression (4) is
incorrect in the case of one particle as well.} The expression (4)
derived from the Maxwell equations and Lorentz equations contradicts
them. It does not present the energy conservation law \cite {bes1},
\cite{bes2}.

The error in the presented proof consists of the unification of the
physically inconsistent Maxwell and Lorentz equations in one system.
According to Lorentz equation the total energy of a charged particle is
included in the term $\varepsilon _i = m_ic^2\gamma _i$ of the Eq(4).
At the same time according to the Maxwell equations a part of the
energy of the particle of the electromagnetic origin will appear in the
term $\varepsilon _i^{em}$ of the same equation. Just this fact leads
to the logic error in the proof of the energy conservation law and
because of which the equation (4) conflicts with the initial equations.
Now we will illustrate this conclusion by the next example.

\subsection {Example }

An immovable large conducting sphere with a charge $q$ and a particle
with a charge $e$ and mass $m$ are separated by a distance $a$. At some
moment the particle start to move and, being accelerated, leaves to
infinity. Let us compare the energy of the particle calculated from the
law of the energy conservation (4) and from the equations of motion.

Let us write down the expression (4) for an initial and a final states
of the system and equate received expressions. Then we will determine
the kinetic energy of the particle

              \begin {equation} 
            T = mc^2 (\gamma - 1) = {eq\over a} - \varepsilon ^ {em}
            _ {rad} - (\varepsilon ^ {em} _ {e} - \varepsilon ^ {em}
            _ {e0}), \end{equation}
where $eq/a$ is the initial potential energy of the particle,
$\varepsilon _ {rad} ^ {em} $ the independent energy of an
electromagnetic field radiated by the particle, $\varepsilon _e^ {em} $
and $\varepsilon ^ {em} _ {e0} $ the inertial energy of the moving
particle and the particle at rest respectively.

Calculation of a kinetic energy of the particle by the solution of the
equation of motion of the particle in the given field of the charge
$q$ will lead to an expression

             \begin {equation} 
             T = {eq\over a} - \varepsilon ^ {em} _ {rad}.  \end {equation}

As was to be expected, an extra term in (5) is equal to a difference
between inertial energies of electromagnetic fields of accelerated and
motionless particles. For the spherically symmetric distribution of
particle's charge the value $\varepsilon ^ {em} _ {e} - \varepsilon ^
{em} _ {e0} = \varepsilon ^ {em} _ {e0} [\gamma (1 + v^2/3c^2) - 1] \ne
0$ \cite {ivan}, \cite{barut}. We can see that the solution (5) of the
problem based on the Eq(4) is incorrect. It differs from the correct
solution (6) by the difference between the field energies of moving
particle and the particle at rest. It's just what has to be expected
when we use the Eq(4).

\subsection{Discussion}

It is not difficult to understand the reasons according to which the
logic error was not exposed for a so prolonged time. The way of the
electrodynamics development was the next. According to the special
theory of relativity the energy and momentum of particles should have
certain transformation properties regardless to theirs nature. The
relativistic mass is a coefficient between vectors of the momentum and
velocity of particles. In this case the Newton's second law and the
Lorentz force govern the dynamics of particles. The joint solution of
the equations of motion of fields and particles reduces to the
expression (4) which is treated as the energy conservation law. At that
it is postulated that the value $w$ is the energy density of the
electromagnetic field. After this postulate was accepted the authors do
not notice that at the same time this postulate leads to the appearance
of the additional unwanted energy of particles of the electromagnetic
origin which is hidden in the total field energy of the system and
violate the sense of the received equation. The nature of mass, energy
and momentum of particles are discussed later, after the Eq(4) is
received and interpreted. They are discussed after the electromagnetic
fields and the corresponding to these fields energy and momentum of the
electromagnetic origin created by a uniformly moving particles are
calculated. It turned out that these values have not correct
transformation properties following from the special theory of
relativity\footnote{In the case of the spherically symmetrical
particles besides the correct factor $\gamma $ there are the incorrect
factors $ (1 + v^2/3c^2) $ in energy and (4/3) in the momentum of
particles \cite {ivan}. In the case of a the non-symmetrical particles
these factors are more complicated and depend on the orientation of the
particles to the direction of theirs velocity. These factors are
changing in time when the particles are in state of rotation. The
corresponding change in time of the energy and momentum of the
particles of the electromagnetic origin caused by theirs rotation can
be compensated only by the fields of non-electromagnetic origin but
this compensation does not work in the case of the short-range
fields.}. In order to give the correct transformation properties and
with the purpose of keeping of the particle charge in equilibrium the
attraction fields of non-electromagnetic origin are introduced. It is
postulated that the energy and momentum of these fields have wrong
transformation properties of such a form that the sum of the energies
of the fields of the electromagnetic and non-electromagnetic origin are
reduced to experimentally observable values of energy and momentum of
particles\footnote {In the case of a non-uniform motion the
concepts of the energy and momentum of particles of the electromagnetic
origin were not discussed.  Non-obviously they are taken equal to the
appropriate values for the particles moving uniformly with the same
velocity.}. Again, after these in word only assumptions there is no
discussion of the necessity of taking into account the presence of the
fields of the non-electromagnetic energy in the equations of motion of
particles and fields and there is no connection of this discussion with
any reference to the conservation law (4) and its revision. After all,
the equations are not changed. Observable mass $m_i$ accepted in word
only. It is finite even in the case of point particles. Both the
inertial and radiation electric fields of the electromagnetic origin
that is the first ($\sim \dot {\vec v}$) and higher terms of the
Abraham-Lorentz self-force \cite{jack} was kept and work but the
corresponding forces of the non-electromagnetic origin were not
introduced and that is why they do not compensate the corresponding
part of the energy and momenta of the field term of the Eq(4)
(specifically, among their number the field energy term $\varepsilon
^{em}_{e} - \varepsilon ^{em}_{e0}$ in the example 1).

Unfortunately the laws of conservation, as a rule, were proved only to
emphasize the consistency of the electrodynamics, its perfection.  They
were not used on practice since on the basis of the laws it is possible
to solve a small number of simple problems not representing practical
interest\footnote {It is possible to point out only on the paper
\cite{bolot} and comment to this paper \cite{bes3} and close questions
connected, for example, with renormalization of mass in a classical
electrodynamics (see [1], $\S$ 37, $\S$ 65, Problem 1 ).  The term
"renormalization procedure" was appeared first in the quantum theory
where the energy field term similar to the field term in (4) was
introduced in the Hamiltonian and where the removal of divergences was
done by veiling the problem by some artifices \cite{markov}. At that
there was no reference on the validity of the energy conservation law
(4) after such artifices were introduced."}.  Therefore the error in
the proof, which would be possible to establish by a comparison of the
solutions following from the laws of conservation and from the
equations of motion, on particular examples was not discovered.

We have shown that the energy conservation law (4) in the case when the
kinetic energy of particles is changed lead to the solution of the
problems which differ from the solutions derived from the equations of
motion of particles in the electromagnetic fields. The considered
example and the paper \cite{bolot} have demonstrated this result. At
the same time in the case when the problem deal with the initial and
final states of the system we can remove the kinetic energy of the
particle of the electromagnetic origin (similar to the last term in the
expression (5)) by hands and such a way to come to a correct solution.
But such solution of the problem (renormalization by hands) means that
the Maxwell-Lorentz electrodynamics is not correct. Moreover this
solution is not universal. The last affirmation was demonstrated in
\cite{bes2} where the problem of two identical charged particles was
considered. In this problem the particles were brought close to each
other with equal constant velocities by extraneous forces along an axis
$x$ up to a distance $a_1$. Then the extraneous forces are switched off
and particles where being decelerated continued to be brought closer by
inertia until they were stopped on a distance $a_2 < a_1$. After the
stop of particles the extraneous forces where switched on again to keep
particles in the state of rest. It was shown that contrary to the law
of conservation of energy the energy of the considered system in the
final state is higher than the energy spent by extraneous forces on
acceleration of particles and their further bringing closer. The idea
of this example was the next. The electrical field of a uniformly
moving charged particle is flattened in the direction of motion such a
way that on the axis of motion at a distance $a$ its value $E_{||}$ is
$\gamma ^2$ times less then the electrostatic field of the particle at
rest being at the same distance from the observation point \cite
{landau}. As the repulsive forces between moving particles $eE_ {||}
^{mov} $ are weakened $\gamma ^2$ times in comparison with a static
case then on a principle of bringing close the particles to each other
with relativistic velocities and subsequent separation them under
non-relativistic velocities one could construct the perpetum
mobile\footnote{In the approximation $(v/c)^2$ this problem can
be solved without introduction of the extraneous forces (Usual
solution of the equations of motion or Darwin Lagrangian can be used
\cite{landau}, \cite{jack}).  The result will be the same.}.
This problem means that either possible form of fields of
non-electromagnetic origin must be restricted or Maxwell-Lorentz
equations must be changed.

                       \subsection {Remarks}

\hskip 6mm
1. The electric and magnetic field strengths in the expressions for the
Lorentz force, $\varepsilon ^{em}$ and $(\vec \jmath\cdot \vec
E)$ according to Maxwell equations include both the external fields and
self-fields produced by charged particles. That is why the radiation
reaction force was taken into account in (2), (4). The energy
conservation law was derived without any assumptions on the value of
the external fields or distances. This statement and the discussion of
the particle nature we can see e.g. in \cite{landau} in the form:"One
can raise the question of how the electrodynamics, which satisfies the
law of conservation of energy, can lead to the absurd result that a
free charge increases its energy without limit.  Actually the roots of
this difficulty lie in the earlier remarks (Section 5-2) concerning the
infinite electromagnetic "intrinsic mass" of elementary particles.
When in the equations of motion we write a finite mass for the charge,
then in doing this we essentially assign to it formally an infinite
negative "mass" of non-electromagnetic origin, which together with the
electromagnetic mass should result in a finite mass for the particle.
Since, however, the subtraction of one infinity from another is not
entirely correct mathematical operation, this leads to a series of
further difficulties, among which is the one mentioned here". It means
that the authors consider the statement "the electrodynamics, which
satisfies the law of conservation of energy" as the fundamental law of
the electrodynamics. They explain the difficulties of the "runaway"
solutions by the not entirely correct mathematical operation connected
with pointlike dimensions of particles.

2. In the textbooks the energy conservation law has the form
$d\varepsilon _{\Sigma}/dt = 0$ that is expressed through the change of
the total energy. Maybe this fact was the reason of the fact that the
divergence of the value $\varepsilon _{\Sigma}$ was not discussed with
the reference to the energy conservation law in the case of the
pointlike particles in spite of such particles were considered in
\cite{landau} and other textbooks.

3. In the electrodynamics of Maxwell-Lorentz there could not be a model
of particles with pure electromagnetic nature of mass. Differently all
energy would have electromagnetic nature and the first term in (4)
should be absent. On the other hand the energy of charged particle
$\varepsilon _i$ cannot have pure non-electromagnetic nature since in
the case, for example, of one particle the energy of extraneous forces
applied to a particle will be transformed not only to the value
$\varepsilon _i$ and to the energy of the emitted radiation but also to
the electromagnetic self-energy $\varepsilon ^ {em} $ of the particle.

4. For the case of one uniformly moving particle the value
$\varepsilon _ {\Sigma} $ depends on a velocity by the law which
differs from the relativistic law. In this case the change of the total
energy of the particle determined by the Eq(4) $\varepsilon _{\Sigma}
\ne  \gamma \varepsilon _{\Sigma \,0}$ as the correct transformation
properties has the first but has not the second term of this expression
representing the electromagnetic self-energy of the particle \cite
{ivan}. The similar statement is valid for the beam of charged
particles where the situation is intensified in addition by the
circumstance that besides the electromagnetic self-energy of particles
the second term in (4) will include the electromagnetic energy of
interactions of particles which have incorrect and more complicated
transformation properties as well.

             \section {Conclusion}

In electrodynamics there are many "open" or "perpetual" problems such
as the problem of the self-energy and momentum of particles, the nature
of the particle's mass, the problem of the runaway solutions. There is
a spectrum of opinions concerning the importance and the ways of
finding of the answers on these questions \cite{bes2}.  Unfortunately
the efforts of the majority of the authors are directed to avoid
similar questions but not to solve them (see e.g. \cite{landau},
\cite{jack}, \cite{ivan}, \cite{markov}, \cite{ginzb}, \cite{feynman}).
In addition they base themselves on the laws of conservation ostensibly
following from the electrodynamics in the most general case and
presenting electrodynamics as the consistent theory. In such stating
the arising questions by their opinion do not have a physical subject
of principle and the difficulties in their solution are on the whole in
the field of the mathematicians. Of cause there is an opinion that the
classical electrodynamics of Maxwell-Lorentz must be changed \cite
{feynman}. Another Non-Maxwellian Electrodynamics were suggested but
none was survived \cite {feynman}\footnote{Nothing changed in the
electrodynamics up to this time.}. All open questions are now remain
unsolved.

It is shown in the present work that contrary to the universal opinion
the relation (4) does not express the energy conservation law in
electrodynamics of Maxwell-Lorentz. The error in the treatment of this
expression is the consequence of the insufficiently precise definitions
of the basic concepts of the theory and its logically inconsistent
construction. It follows that in the process of any discussion of the
existing difficulties of the classical electrodynamics it is impossible
to refer to the energy conservation law in the form, which was done,
for example, in the textbook \cite {landau}, and the paper \cite
{ginzb}. The same confirmation is valid for the linear and angular
momentum conservation laws as well.

We would like to have classical electrodynamics on a level with
classical mechanics in the form of consistent theory. Let the theory
doesn't agree with experiment which require the quantum mechanical
generalization. The generalization of the consistent classical theory
will lead to the consistent quantum theory. There are difficulties
associated with the ideas of Maxwell's theory which are not solved by
and not directly associated with quantum mechanics \cite{feynman},
\cite {schwinger}. This difficulties including the problem of the
violation of the energy conservation law for a system of particles and
electromagnetic fields must be widely presented in the textbooks
devoted to the foundations of classical electrodynamics. We hope for
the more comprehensive analysis and further developments of the
classical and quantum electrodynamics.

\section {Appendix}

The electromagnetic field in vacuum is described by the Maxwell
equations

          \begin {equation} 
           rot \vec E = - {\partial \vec H\over \partial t},
          \end {equation}
          \begin {equation} 
           div \vec H =0,
           \end {equation}
           \begin {equation} 
            rot \vec H = {4\pi\over c} \vec j + {1\over c} {\partial \vec E\over
           \partial t}
           \end {equation}
           \begin {equation} 
            div \vec E = 4\pi \rho
           \end {equation}

The typical proof of the law of conservation of energy in
electrodynamics is derived according to the following scheme \cite
{landau}.

Let us multiply both parts of the equation (7) by $\vec H$ and both
parts of the equation (9) by $\vec E$ and subtract the received
equations term by term

           \begin {equation} 
          {1\over c} (\vec E {\partial \vec E\over \partial t}) +
          {1\over c} (\vec H {\partial \vec H\over \partial t}) =
          - {4\pi\over c} \vec j \vec E - (\vec H rot \vec E - \vec
          E rot \vec H). \end {equation}

Using the known formula of the vector analysis $div [\vec a\vec b] =
\vec b rot\,\vec a - \vec a rot\,\vec b$ we rewrite this relation in
the form
            \begin {equation} 
          {\partial \over \partial t} {| \vec E | ^2 + | \vec H | ^2\over 8\pi}
            = - \vec j \vec E - div \vec S, \end {equation}
where $\vec S= (c/4\pi) [\vec E\vec H] $ is the Pointing vector.

Let us integrate (12) through some volume and apply the Gauss theorem
to the second term from the right. Then we will receive the equation
(1). If the system consists of charged particles then the integral
$\int \vec j \vec E dV$ can be written down in the form of a sum
corresponding to all particles of the system of a form $\sum e\vec
v_i \vec E (\vec r_i) = \sum d\varepsilon _i/dt$. In this case the
equation (1) is transformed into the equation

              \begin {equation} 
             {\partial \over \partial t} (\int {| \vec E | ^2 + | \vec
              H | ^2\over 8\pi} dV + \sum _i \varepsilon _i) = - \oint
              \vec S d\vec f.  \end {equation}
which is equivalent to (2).

The value {\Large $\oint $} $\vec S d\vec f$ is a flow of the energy of
the electromagnetic field through a surface limiting the volume. If the
integration is made through the whole volume of the space i.e. if a
surface of integration is withdrawn to infinity then the surface
integral is disappeared (all particles and fields remain in space and
do not go outside of the limits of the surface of integration) \cite
{landau}. In this case the Eq(13) will be transformed into the Eq(4).

We can see that the deduction of the equations (1)-(4), (13) was done
at an arbitrary reference system. It means that the values $w$, $\vec
S$ are described by the same form independently on reference frame. The
vector $\vec S$ is included in the operator $div$ and that is why the
Poynting vector $\vec S$ is determined with the accuracy to a $rot$ of
an arbitrary vector $\vec A$. But such ambiguity of $\vec S$ do not
lead to some significant physical consequence as the value $\oint rot
\vec A d\vec f = 0$. We have presented this remark in detail because of
the erroneous papers appeared where the form of the values $w$ and
$\vec S$ depends on the reference system (see e.g.  the paper
\cite{rohrlich} which unfortunately was cited without any criticism in
some papers (e.g. \cite {schwinger}) and in the textbook \cite {jack}).

\end{document}